# Aharonov-Bohm effect and classical Hamiltonian mechanics

Alexander Ershkovich and Peter Israelevich

Department of Geophysics and Planetary Sciences

Tel Aviv University, Ramat Aviv, Israel

Aharonov-Bohm effect (1959) was always considered to be a quantum mechanics one. For instance, Feynman et al. [1] mentioned that it was implicit from the beginning of quantum mechanics in 1926. One of us, A.E. [2] drew attention to the fact that Hamilton-Jacobi equation which governs the classical mechanics explicitly depends on the potential **A** of the magnetic field, like Schroedinger equation, and hence **A** should be a physical reality (rather than a mathematical artifice) in classical physics as well. This is quite natural as the classical mechanics follows from Hamilton principle of least action, and the action $S$ contains the term

$$S_{in} = q \int \mathbf{A} d\mathbf{r} \qquad (1)$$

which is indicative of the electric charge $q$ interaction with the magnetic field $\mathbf{B} = \nabla \times \mathbf{A}$ (or more correctly, with the field **A**).

Let us show that Aharonov-Bohm effect may be derived from the Hamilton-Jacobi equation using de Broglie idea (1924) on wave-particle duality.

Hamilton-Jacobi equation for the action $S$ is

$$\frac{\partial S}{\partial t} + H = 0 \qquad (2)$$

where $H = \frac{1}{2m}\left(\frac{\partial S}{\partial \mathbf{r}} - q\mathbf{A}\right)^2$ is the Hamiltonian. The phase $\Psi$ of the de Broglie wave obeys the eikonal equation in geometrical optics

$$\frac{\partial \Psi}{\partial t} + \omega = 0 \qquad (3)$$

With the energy $H = \hbar\omega$, and the generalized momentum $\mathbf{P} = \hbar\mathbf{k}$, equation (3) differs from (2) only by the factor $\hbar = H/\omega = S/\Psi$. Thus the phase shift $\Psi_s$ accumulated due to particle field-interaction is $\Psi_s = S_{in}/\hbar$. Using equation (1) we arrived at the Aharonov-Bohm effect:

$$\Psi_s = \frac{q}{\hbar}\int \mathbf{A} d\mathbf{r} \qquad (4)$$

Schroedinger equation was not required. It means that Aharonov-Bohm effect, strictly speaking, is not a quantum mechanics effect. It might have been predicted well before the quantum mechanics creation by Heisenberg and Schroedinger.

Alexander Ershkovich

(alexer@post.tau.ac.il)

Peter Israelevich

(peteri@post.tau.ac.il)

Tel Aviv University